\documentclass[12pt,a4paper,final]{iopart}
\pdfoutput=1
\usepackage{graphicx}
\usepackage{cite}
\usepackage[breaklinks=true,colorlinks=true,linkcolor=blue,urlcolor=blue,citecolor=blue]{hyperref}
\usepackage{amsfonts, amsmath, dsfont, amssymb}
\def\ba{\begin{equation}\begin{array}{c}}
\def\ea{\end{array}\end{equation}}
\def\be{\ba\displaystyle}
\def\ee{\ea}
\usepackage{blkarray}
\def\Big#1{\makebox(0,0){\Large#1}}

\begin{document}

\title{Soliton splitting in quenched classical integrable systems}

\author{O. Gamayun}

\address{Instituut-Lorentz, Universiteit Leiden, P.O. Box 9506, 2300 RA Leiden, The Netherlands}

\ead{gamayun@lorentz.leidenuniv.nl}

\author{M. Semenyakin}

\address{Department of Physics, Taras Shevchenko National University of Kyiv 64/13, Volodymyrska Street, Kyiv 01601, Ukraine, and}

\address{Bogolyubov Institute for Theoretical Physics, 14-b Metrolohichna str., Kyiv 03680, Ukraine}

\ead{semenyakinms@gmail.com}

\begin{abstract}

We take a soliton solution of a classical non-linear integrable equation and quench (suddenly change) its non-linearity parameter. 
For that we multiply the amplitude or the width of a soliton by a numerical factor $\eta$ and take the 
obtained profile as a new initial condition. 
We find the values of $\eta$ at which the post-quench solution consists of only a finite number of solitons. 
The parameters of these solitons are found explicitly.
Our approach is based 
on solving the direct scattering problem analytically. We demonstrate how it works for Kortewig-de-Vries, sine-Gordon and non-linear Schr\"odinger integrable equations.

\end{abstract}

\section{Introduction}

One of the most remarkable phenomena of the nonlinear dynamics is the existence of a soliton --- a particle-like solitary wave-packet propagating without changing its shape \cite{Rajaraman}.
Solitons have been observed in a large variety of systems. They appear as optical pulses in semiconducting waveguides \cite{Allan:91,PhysRevLett.81.3383} and optical fibers \cite{optic,5,PhysRevLett.60.29},  waves in shallow fluids \cite{kdvor,PhysRevLett.64.1518}, standing waves in mechanical systems \cite{PhysRevLett.68.1730}, dissipative nonlinear waves in complex plasmas \cite{PhysRevLett.102.135002}, excitations in molecular systems \cite{Dav}, magnetic flux quanta in Josephson junctions \cite{jj} and density excitations in cold atomic systems \cite{PhysRevLett.83.5198,Becker,PhysRevLett.101.130401,Stellmer2008,Mendonca2013,NONLINEAR,Frantzeskakis2010}.

Changing parameters in non-linear equations leads to interesting dynamics and pattern formations \cite{Konotop2,Konotop1}. 
This can be used, in particular, to generate solitons. A promising class of systems for such kind of 
changes are cold atomic systems where control and manipulation of the microscopic parameters of the Hamiltonian have achieved remarkable results \cite{Greiner,Kinoshita2006,Schmiedmayer2007,bloch2008many}. 
Significant efforts have been devoted to explore systems driven out of equilibrium by an instantaneous change of one or several parameters \cite{Haller2009,Gring,Trotzky,Ferrari}.
 This process goes by the name of global quench. Post-quench relaxation of the system  and the properties of its steady state
were investigated \cite{Dj,RevModPhys.83.863,JEisert}. It was shown, in particular, that relaxation in the case of integrable system behaves in a rather unusual way due to the presence of an infinite number of the conserved charges 
\cite{PhysRevA.74.053616,cauxEssler,PhysRevLett.100.030602,PhysRevLett.106.227203,1742-5468-2012-07-P07016,PhysRevB.87.245107,1367-2630-12-5-055015,PhysRevLett.109.175301,PhysRevLett.109.247206,PhysRevLett.106.140405,PhysRevLett.111.100401,1742-5468-2013-07-P07003,1742-5468-2013-07-P07012,PhysRevB.89.125101,PhysRevLett.113.117202,PhysRevLett.113.117203,PhysRevA.89.013609,PhysRevA.89.033601,1742-5468-2014-7-P07024,PhysRevA.91.051602,PhysRevLett.115.157201}.

A quasiclassical description of Bose-Einstein condensate (BEC) of cold atoms is provided by Gross-Pitaevskii equation \cite{Pitaevskii2003}.  We refer to the Gross-Pitaevski equation in one spatial dimension (1D) as a non-linear Schr\"odinger equation (NLS). This equation is  integrable and solitons are its most famous solutions \cite{Faddeev,Babelon,ZMNP}. 
Recently, the problem of quenching the soliton profile in 1D BEC at finite density has been investigated in Refs. \cite{PhysRevA.91.031605,Franchini}. In Ref. \cite{PhysRevA.91.031605} it was found that if the coupling in BEC is quenched 
in such a way that the speed of sound in the condensate is increased by an integer number $\eta$ then 
the initial soliton splits exactly into the $2\eta -1$ solitons. These quenched solitons can be divided by the two groups of $\eta$ and $\eta-1$ solitons that are moving in the same and in the opposite direction of the initial soliton, correspondingly. 
For a short period of time these groups look like two bumps moving in the opposite directions. This short time dynamics of such structures was investigated in Ref. \cite{Franchini}. The shape and velocity of each bump were found for abritrary $\eta$
and its relation to a solution of the Kortewig-de-Vries equation (KdV) was discussed.
The similar question for the bright soliton was explored in Ref. \cite{Miles}. 

In this paper we extend the approach of Ref. \cite{PhysRevA.91.031605} 
to investigate quenches in arbitrary classical integrable systems. We employ the fact that the general solution of such systems can be obtained by the
inverse scattering transformation (IST) method \cite{Faddeev,Babelon,ZMNP}. In this method the non-linear equation is replaced by the consistency condition for the set of linear differential equations, one of which can be interpreted as a scattering problem on a given potential. It turns out that the time evolution of the scattering data (the transfer matrix)
is quite simple due to the integrability. Thus, solving the Cauchy problem can be done in two steps. The first one, called the 
the direct scattering problem, is a calculation of scattering data on the initial field configuration. 
The second one is recovering the potential from the time evolved scattering data. 
The latter is an extremely difficult task for a generic initial condition as it involves solution of a linear integral equation. However, there are special cases of the solitonic solutions that correspond to the reflectionless potentials (diagonal transfer matrix) and the IST reduces to the linear algebraic equations. 
In our approach, we solve analytically direct scattering problem for a one-soliton profile quenched by an arbitrary factor $\eta$.
We find all values of $\eta$ whose potential is reflectionless, and thus corresponds to the solitonic solution where the subsequent time dynamics can be found explicitly. 

The paper is organized as follows. In Sec. \ref{Lemma} we describe general transfer matrix for the scattering on a quenched soliton profile. In Sec. \ref{KdV} using general transfer matrix we find the conditions for quench to produce solitons in KdV equation, and describe the parameters of these solitons. Briefly, we describe effects of generic $\eta$. In Sec. \ref{NLSA} we present the same analysis for NLS equation at zero density and attractive interaction. In Sec. \ref{SG} 
we analyse sine-Gordon equation and Sec. \ref{NLS} is devoted to the the repulsive NLS equation at finite density.  
Our conclusions are presented in Sec. \ref{Conclusions}.

\section{General transfer matrix}
\label{Lemma}

We treat the solution of an integrable system by the methods described in textbook \cite{Faddeev}. Namely, a non-linear equation is replaced by 
the compatibility condition for a linear system
\be\label{U}
\frac{\partial F}{\partial x} = U(x,t) F\,,
\ee
\be\label{V}
\frac{\partial F}{\partial t} = V(x,t) F\,.
\ee
called zero curvature condition
\be\label{UV}
\frac{\partial U}{\partial t} - \frac{\partial V}{\partial x} + [U,V] = 0\,.
\ee
Here $U(x,t)$ and $V(x,t)$ are $2\times 2$ matrices that are composed of the dynamical variables and act in the two-dimensional auxiliary space. 
For a given initial condition Eq. (\ref{U}) defines a scattering problem. The corresponding scattering data (transfer matrix) evolves in time with respect to Eq. (\ref{V}). Its evolution 
is remarkably simple because of the presence of the condition (\ref{UV}), which in this case is analogous to integrability. 
This way, solving the original problem reduces to finding potential $U(x,t)$ in Eq. (\ref{U}) given a set of scattering data. Such procedure is called the inverse scattering transformation (IST). As we have mentioned in introduction, in the case of reflectionless potential (\ref{U}) IST is basically a linear algebraic problem, which can be easily solved, and the obtained solution is called solitonic. 

In this paper we consider quenched one-solitonic solution as the initial condition. 
We find that it is more convenient to perform change of variables from $x \to z$ such that 
asymptotic regions $x\to\mp\infty$ would correspond to $z\to 0$ and $z\to 1$, respectively.  
It turns out that we can always 
perform a gauge transformation to present the scattering problem (\ref{U}) with our initial condition in a Fuchs form
\be\label{1.1}
\frac{d F}{d z} = \left(\frac{A_0}{z}  + \frac{A_1}{1-z}\right) F\,,
\ee
with constant matrices $A_0$ and $A_1$. 
Moreover, we can always make these matrices degenerate, so they can be expressed as
\be
A_0 = G_0 \left(
 \begin{array}{cc}
        0 & 0 \\
        0 & \lambda_0 \\
      \end{array}
    \right)G_0^{-1},\,\,\,\,\,
A_1 = G_1 \left(
 \begin{array}{cc}
        0 & 0 \\
        0 & \lambda_1 \\
      \end{array}
    \right) G_1^{-1}\,.
\ee
The Jost solutions of Eq. (\ref{1.1}) are defined through their asymptotic behaviour at $z\to 0$ and $z \to 1$, by the constant matrices $C_0$ and $C_1$\footnote{These matrices are uniquely defined by the required analytic dependence of the spectral parameter   \cite{Faddeev}.}, namely
\begin{equation}\label{Jost}
\fl \quad\quad T_0(z\to 0) = G_0 \left(
 \begin{array}{cc}
        1 & 0 \\
        0 & z^{\lambda_0} \\
      \end{array}
    \right)C_0,\quad
    T_1(z\to 1) = G_1 \left(
 \begin{array}{cc}
        1 & 0 \\
        0 & (1-z)^{-\lambda_1} \\
      \end{array}
    \right)C_1\,.	
\end{equation}
All scattering data are contained in the transfer matrix $T$ that connects these two Jost solutions
\be\label{T}
T_0(z)= T_1(z) T.
\ee
The general expression for the transfer matrix $T$ for a given $A_0$, $A_1$, $C_0$ and $C_1$ can be extracted through the exact solution of Eq. (\ref{1.1}).
The latter can be found easily in the basis where $A_0$ is diagonal. Therefore 
substituting $F = G_0 \Phi$ in Eq. (\ref{1.1}), one can immediately read off the general solution 
\be\label{PHI0}
\Phi(z) = \left(
 \begin{array}{cc}
        w_1(a,b,c;z) & \displaystyle \frac{\alpha\beta\lambda_1}{\lambda_0+1}w_2(a,b,c;z) \\
        \displaystyle z \frac{\gamma\delta}{\lambda_0-1}w_1(\tilde{a},\tilde{b},\tilde{c};z) & \displaystyle  z w_2(\tilde{a},\tilde{b},\tilde{c};z) \\
      \end{array}
    \right) C
\ee
where the coefficients are given by the conjugation matrices
\be
G = G_0^{-1}G_1 = \left(
 \begin{array}{cc}
        \alpha & \beta \\
        \gamma & (1+\beta \gamma)/\alpha \\
      \end{array}
    \right) = \left(
 \begin{array}{cc}
        \alpha & \beta \\
        \gamma & \delta \\
      \end{array}
    \right),
\ee
and $w_i(a,b,c,z)$ corresponds to the specific solution of the hypergeometric equation according to \cite{Olver} ({\tt http://dlmf.nist.gov/15.10})
\be
\fl w_1(a,b,c,z) = \,_2F_1(a,b,c;z),\,\,\,w_2(a,b,c,z) = z^{1-c}\,_2F_1(a-c+1,b-c+1,2-c;z).
\ee
The corresponding values of the parameters read as
\be\label{par1}
 a=\tilde{a}-1=\theta^+ ,\,\,\,\,\,\,b=\tilde{b}-1 = \theta^- ,\,\,\,\,\,\,  c = - {\rm tr} A_0,\,\,\,\,\,\, \tilde{c} = 2- {\rm tr} A_0
\ee 
where
\be\label{par2}
\theta^\pm = \frac{{\rm tr}A_1-{\rm tr}A_0}{2} \pm
\frac{1}{2}\sqrt{({\rm tr}A_0+{\rm tr}A_1)^2-{\rm tr}A_0A_1}.
\ee
To express all parameters of the hypergeometric functions through 
the invariants of $A_0$ and $A_1$ we have used the identities
\be
\lambda_0 = {\rm tr} A_0,\,\,\,\,\,\,\lambda_1 = {\rm tr} A_1,\,\,\,\,\,\, (1+\beta\gamma)\lambda_0\lambda_1 = {\rm tr}A_0A_1.
\ee
The matrix $C$ is responsible for the initial conditions. If we fix it in such a way that the corresponding asymptotic of the solution
coincide with the Jost one (\ref{Jost}), then using connectivity formulas for the hypergeometic functions \cite{Olver} and 
relation (\ref{T}) we immediately find the desired transfer matrix

\be \label{transfer}
\fl \quad\quad
 T = C_1^{-1} \left(
 \begin{array}{cc}
        \displaystyle\frac{\Gamma(c)\Gamma(c-a-b)}{\alpha\Gamma(c-a)\Gamma(c-b)} & \displaystyle\frac{\beta \lambda_1\Gamma(2-c)\Gamma(c-a-b)}{(1+\lambda_0)\Gamma(1-a)\Gamma(1-b)} \\
        \displaystyle\frac{\Gamma(c)\Gamma(a+b-c)}{\beta\Gamma(a)\Gamma(b)}&  \displaystyle\frac{\lambda_1 \alpha \Gamma(2-c)\Gamma(a+b-c)}{\Gamma(a-c+1)\Gamma(b-c+1)(1+\lambda_0)}\\
      \end{array}
    \right)
    C_0.
\ee

This is the general expression for the transfer matrix that describes scattering on a soliton like potential. 
Given a specific integrable system one has to transform the scattering problem to the Fuchs form (\ref{1.1})
and find corresponding asymptotics of Jost solutions (\ref{Jost}), after which the transfer matrix (\ref{transfer}) 
comes immediately. We implement this program for various integrable systems in sections \ref{KdV}, \ref{NLSA}, \ref{SG} and  \ref{NLS}.
We focus on the conditions when this matrix is diagonal, which means that the potential is reflectionless. In these cases we find 
specific solitonic form of the solutions.

\section{Kortewig-De Vries equation}
\label{KdV}

Let us first demonstrate how the procedure outlined above works for the famous KdV equation
\begin{equation}\label{kdve}
\frac{\partial u}{\partial t}-6u\frac{\partial u}{\partial x}+\frac{\partial^3 u}{\partial x^3}=0.
\end{equation}
The one-soliton solution of the KdV equation has a profile of a specific form that propagates without changing its shape with time, namely
\be
u(x,t) =  - \frac{2\varkappa^2}{\cosh^{2}(\varkappa (x - 4\varkappa^2 t))} .
\ee
Now assume that the initial condition corresponds to the quenched profile
\be\label{u0}
u(x,0) \to u_{\eta}(x) =  - \frac{2\varkappa^2}{\cosh^{2}(\varkappa x/\eta)}.
\ee
Our main question is how the system evolves starting from this profile for different $\eta$.  Further for simplicity we put $\varkappa=1$. 
Eq. (\ref{kdve}) is integrable by the means of IST and the corresponding direct scattering problem is given by the system \cite{Faddeev}
\be \label{KDV_U_evolution0}
\frac{dF}{dx}
=
\left(
\begin{array}{cc}
\displaystyle\frac{\lambda}{2i} & 1 \\
u_\eta(x) & -\displaystyle\frac{\lambda}{2i}
\end{array}
\right)F.
\ee
The Jost solutions of this equation are determined by their asymptotic behaviour
\be\label{JostKDV}
T_{\pm} \to \left(
 \begin{array}{cc}
        1 & 1 \\
        0 & i\lambda \\
      \end{array}
    \right)\left(
 \begin{array}{cc}
        e^{-i{x\lambda/2}} & 0 \\
        0 & e^{ix\lambda/2} \\
      \end{array}
    \right)\equiv S e^{-i\sigma_3\lambda x/2},\,\,\,\, x \to \pm \infty.
\ee
We can employ gauge transformation
\be\label{GaugeF}
F \to \left(
 \begin{array}{cc}
        0 & e^{-x/\eta + i\lambda x/2}\cosh(x/\eta) \\
        e^{i\lambda x/2} & 0 \\
      \end{array}
    \right) F,
\ee
and use new coordinate $z = (1+\tanh(x/\eta))/2$ to present (\ref{KDV_U_evolution0}) in a Fuchs form, namely
\be
\label{KDV_U_evolution1}
\fl \quad \frac{dF}{dz}
=
\frac{1}{z}\left(
\begin{array}{cc}
0 & -2\eta   \\
0 & 1-i\lambda\eta/2
\end{array}
\right)F +\frac{1}{1-z}\left(
\begin{array}{cc}
0 & 0 \\
\eta & -i\lambda\eta/2
\end{array}
\right)F \equiv \left(\frac{A_0}{z}+\frac{A_1}{1-z}\right)F.
\ee
Matrices $C_0$ and $C_1$ that correspond to Jost solutions can be found by comparing the general asymptotic behaviour (\ref{Jost}) with the corresponding one for KdV (Eq. (\ref{JostKDV})). Here we provide detailed derivation, while for the integrable systems in other sections of this paper we just present the final answers. 

At $x\to+\infty$, i.e. $z=1-e^{-2x/\eta}\to 1$, the asymptotic for $T_+$ from Eq. 
(\ref{JostKDV}), combined with the gauge transformation (\ref{GaugeF}) should be matched with the general asymptotic (\ref{Jost}). Namely,
\be\fl
 S \left(
 \begin{array}{cc}
        (1-z)^{i\lambda\eta/4} & 0 \\
        0 & (1-z)^{-i\lambda\eta/4}  \\
      \end{array}
    \right) =(1-z)^{-i\lambda\eta/4}
\left(
 \begin{array}{cc}
        0 & 1/2 \\
        1 & 0  \\
      \end{array}
    \right) G_1\left(
 \begin{array}{cc}
        1 & 0 \\
        0 & (1-z)^{i\lambda\eta/2} \\
      \end{array}
    \right)C_1.
\ee
The matrix $G_1$ that diagonalizes $A_1$ is chosen to be
\be
G_1= \left(
\begin{array}{cc}
 1 & 0 \\
 -2 i/\lambda & 1
\end{array}
\right).
\ee 
which gives 
\be
C_1 = \left(
 \begin{array}{cc}
        0 & i \lambda \\
        2 & 0 \\
      \end{array}
    \right).
\ee
Analogous procedure for $C_0$ requires slight modification because of the specific form of the gauge transformation. Namely,
to find matching matrix at $x\to-\infty$ and $z= e^{2x/\eta}\to 0$ we have to take into account the next to leading order expansion terms, 
which can be easily done with the help of the exact solution (\ref{PHI0})
\be
\fl S\left(
 \begin{array}{cc}
        z^{-i\lambda\eta/2} & 0 \\
        0 & 1 \\
      \end{array}
    \right) = 
\left(
 \begin{array}{cc}
        0 & 1/(2z) \\
        1 & 0 \\
      \end{array}
    \right)G_0\left[\left(
 \begin{array}{cc}
        1 & 0 \\
        0 & z^{1-i\lambda\eta/2} \\
      \end{array}
    \right)
+     z\left(
 \begin{array}{cc}
        -\beta\gamma & \frac{\alpha\beta}{\lambda_0+1}z^{1-i\lambda\eta/2} \\
        \frac{\gamma\delta}{\lambda_0-1} & \alpha\delta z^{1-i\lambda\eta/2} \\
      \end{array}
    \right)
    \right]C_0.
\ee
This relation implies that $C_0 = C_1$. 

With given $A_0$, $A_1$, $C_0$, $C_1$ the transfer matrix (\ref{transfer}) takes the following form
\be
T = \left(
\begin{array}{cc}
a(\lambda) & b(\lambda) \\
\bar{b}(\lambda) & \bar{a}(\lambda)
\end{array}
\right)
\ee
with 
\be
a(\lambda) = \frac{\Gamma \left(-\frac{i \lambda  \eta }{2} \right) \Gamma \left(1-\frac{i \lambda  \eta }{2}\right)}{\Gamma \left(\frac{1}{2} \left(1-i \lambda  \eta -\sqrt{1+8 \eta ^2}\right)\right) \Gamma \left(\frac{1}{2} \left(1-i \lambda  \eta +\sqrt{1+8 \eta ^2}\right)\right)},
\ee
and
\be
b(\lambda) = i\frac{\cos\left(\frac{\pi}{2}   \sqrt{1+8\eta ^2}\right)}{\sinh\left(\frac{\pi  \lambda  \eta }{2}\right)}
\ee
The solitonic case corresponds to $b(\lambda)=0$, which gives
\be
\sqrt{1+8 \eta ^2} = 1+2n\,\,\,\,\, n \in \mathds{Z}.
\ee
This gives 
\be\label{etaKdV}
\eta^2= \frac{n(n+1)}{2}.
\ee
Further, without loss of generality we can assume that $n\ge 1$.
Then the diagonal component of the transfer matrix reads
\be\label{KdVa}
\fl \quad a(\lambda) = \frac{\Gamma \left(-\frac{i \lambda  \eta }{2} \right) \Gamma \left(1-\frac{i \lambda  \eta }{2}\right)}{\Gamma \left(\frac{-i \lambda  \eta}{2}  -n\right)\Gamma \left(\frac{-i \lambda  \eta}{2} +1 +n\right)}   = \prod\limits_{k=1}^n \frac{\lambda-2ik/\eta}{\lambda+2ik/\eta}\equiv \prod\limits_{k=1}^n \frac{\lambda-i\varkappa_k}{\lambda+i\varkappa_k}\,.
\ee
Such a form corresponds to the $n$-solitonic solution. This way, we have found that if the quenching parameter satisfies the condition (\ref{etaKdV}),
after a certain amount of time quenched soliton profile splits exactly into $n$ solitons. 

The general solution of the IST for the reflectionless potential with coefficient $a(\lambda)$ in the above form is given by 
 \cite{ZMNP}
\be\label{uuu}
u(x,t) = -2\; \partial_x^2\ln\det A(x,t)
\ee
where $A(x,t)$ is an $n\times n$ matrix with elements
\be
A_{jk}=\delta _{jk}+\frac{\beta_j}{\varkappa_j+\varkappa_k} e^{-(\varkappa_j+\varkappa_k)x+8 \varkappa_j^3 t}\,.
\ee
Parameters $\beta_j$ play a role of initial conditions that in our case must be chosen to satisfy Eq. (\ref{u0}).
This can be done either by direct matching of the general solution with the initial condition or by taking careful limit 
$\beta_j \sim \lim\limits_{\lambda\to i\varkappa_j }b(\lambda)/(i\partial_\lambda a(\lambda))$ \cite{ZMNP}. The final result reads as 
\be\label{uuu1}
A_{jk}=\delta _{jk}+\frac{(n+j)!}{(n-j)!}\frac{1}{(j!)^2}\frac{j}{j+k} e^{-(\varkappa_j+\varkappa_k)x+8 \varkappa_j^3 t},\quad \varkappa_j
=\frac{2 j}{\eta}\,.
\ee

The time evolution's snapshots for $n=2$ and $n=3$ are shown in Fig. (\ref{KDV_2}) and Fig. (\ref{KDV_3}), respectively.
\begin{center}
\begin{figure}[th]
\includegraphics[width=1\textwidth]{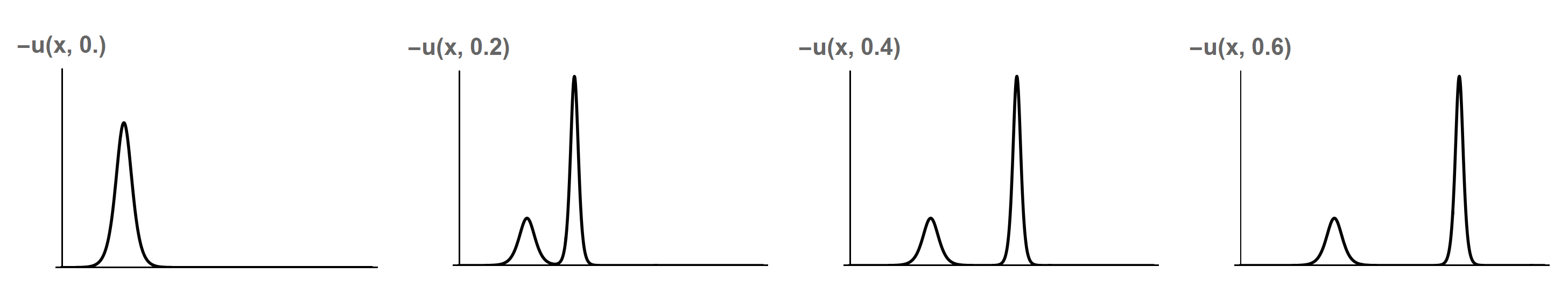}
\caption{Snapshots of the solution of the KdV equation $u(x,t)$ from the quenched one-soliton profile (\ref{u0}) with $\eta=\sqrt{3}$ ($n=2$) calculated from the exact expression Eqs. (\ref{uuu}),(\ref{uuu1})}
\label{KDV_2}
\end{figure}
\begin{figure}[th]
\includegraphics[width=1\textwidth]{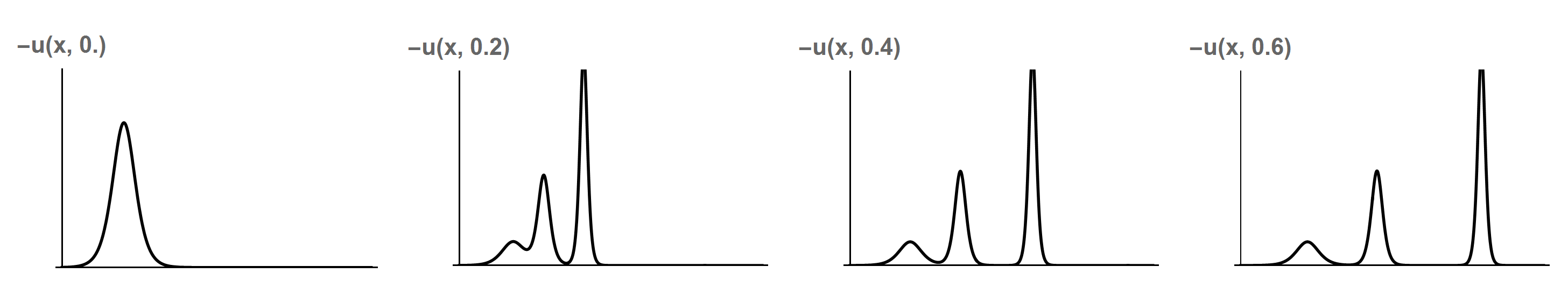}
\caption{Snapshots of the solution of the KdV equation $u(x,t)$ from the quenched one-soliton profile (\ref{u0}) with $\eta=\sqrt{6}$ ($n=3$) calculated from the exact expression Eqs. (\ref{uuu}),(\ref{uuu1})}
\label{KDV_3}
\end{figure}
\end{center}

Integrability imposes the existence of an infinite set of integrals of motion. 
They can be presented as some integrals of local polynomials in field variables. 
Therefore calculating these integrals on the initial profile and 
comparing with those that corresponds to the solitons, one can, in principle, guess under which condition 
the initial profile is a solitonic one. Unfortunately, the straightforward implementation of this task 
can be rather daunting as the explicit form of higher integrals of motion is rather complicated and unknown. 
On the other side, knowing the exact answer (\ref{KdVa}), we can immediately find the value of all integrals of motion
evaluated at the initial quenched soliton profile. 

The integrals of motion can be presented
as integrals of local polynomial densities
\be
I_{n} = \int\limits_{-\infty}^\infty \chi_{n}(x) dx
\ee
that are determined from the recurrence relation 
\be
\chi_{n+1}(x)=\frac{\partial\chi_{n}(x)}{\partial x}+\sum\limits_{k=1}^{n-1}\chi_k(x)\chi_{n-k}(x)\,,
\ee
\be
\chi_{1}=-u(x)\,.
\ee
All even densities are full derivatives, so they give zero integrals of motion. 
The values of first non-trivial integrals which correspond to the initial condition ($\ref{u0}$) are
\be
I_1 = 4\eta,\,\,\,\,\, I_3= \frac{16\eta}{3},\,\,\,\,\, I_5= \frac{64(4\eta^2-1)}{15\eta}\,.
\ee
An alternative way to get the integrals of motion is by using the asymptotic expansion of 
$\ln a(\lambda)$, which is the generator of the local integrals of motion
\be\label{AAAA}
\ln a(\lambda) = 1 + \sum\limits_{j=1}^\infty \frac{I_j}{(i\lambda)^j}+ O(|\lambda|^{-\infty})\,.
\ee 
This way, using the exact expression (\ref{KdVa}) we can easily deduce the general expression for the integrals of motion evaluated at the 
field configuration (\ref{u0}), namely
\be
I_j =  \sum\limits_{k=1}^n \frac{(2k/\eta)^j}{j}\left(1-(-1)^j\right)\,,
\ee
The asymptotic series (\ref{AAAA}), considered as a generating function for the integrals $\{I_j\}$ can be replaced by 
a simpler one, by means of a Borel transformation with a slight readjusting of indices.  Namely, let us introduce function
\be
S(t) = \sum\limits_{j=1}^\infty \frac{I_jt^j}{(j-1)!}.
\ee
Changing the summation order inside each $I_j$ we can easily get
\be
S(t) =\frac{\cosh (t\sqrt{1+8\eta^2}/\eta)-\cosh(t/\eta)}{\sinh (t/\eta)}\,.
\ee
So this "new" generating function is expressed in terms of the elementary functions.
Moreover, one can easily understand that $S(t)$ produces values for the integrals of motion not only for the reflectionless potentials but for generic $\eta$ as well.

\begin{center}
\begin{figure}[th]
\includegraphics[width=1\textwidth]{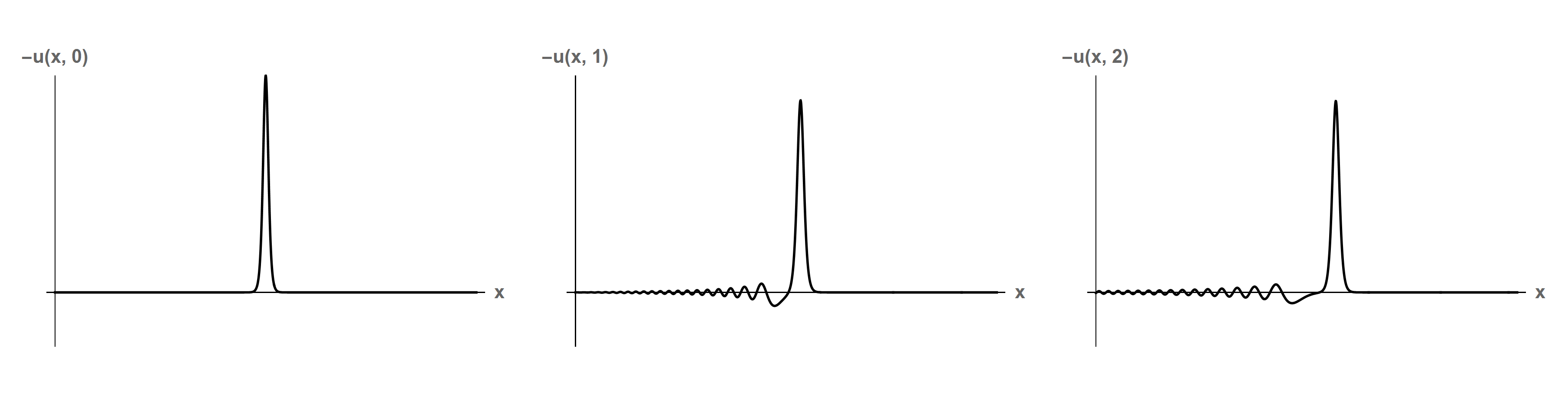}
\caption{Numerically calculated post-quench dynamics for KdV equation from the initial condition (\ref{u0}) with $\eta=0.84$.}
\label{ripples}
\end{figure}
\end{center}

Finally, 
let us comment on what happens if the parameter $n$ determined for the condition (\ref{etaKdV}) is not an integer. 
Using the general considerations \cite{ZMNP} and an exact asymptotic (see Ref. \cite{Grunert} and references therein)
one can argue that in this case the post-quench profile contains 
a radiation part in a form of small oscillatory ripples in addition to $[n]+1$ solitons, with $[n]$ being an integer part of $n$.
For instance, for $0<n<1$ we have one soliton travelling on the oscillating background. Such a solution can be obtained numerically
and an exemplary dynamics is shown in Fig. \ref{ripples}.

\section{Non-linear Schr\"odinger equation. Attractive case}
\label{NLSA}

Our next example of an integrable system with two-dimensional auxiliary space is the non-linear Schr\"odinger equation (NLS) for a rapidly decreasing boundary condition.
It is convenient to fix the following form of this equation
\be\label{NLSa}
i\frac{\partial \psi}{\partial t} = - \frac{\partial^2 \psi}{\partial x^2} + 2\kappa |\psi|^2 \psi\,,\,\,\,\, \psi(x,t) \to 0,\,\,\,\,\,\,\, |x| \to \infty.
\ee
The attractive case corresponds to negative coupling constant $\kappa = -|\kappa|<0$. NLS is a universal way to describe the evolution of wave envelopes in a weakly interacting non-linear medium. It has many applications in non-linear optics, plasma physics, and ultra-cold atom systems and is the main theoretical tool to describe bright solitons. 

\begin{center}
\begin{figure}[th]
\includegraphics[width=1\textwidth]{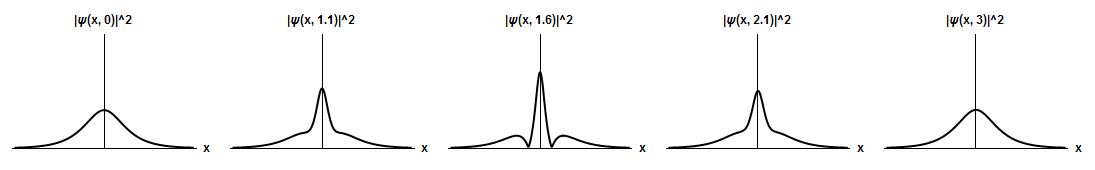}
\caption{Snapshots of the solution of the NLS equation $\psi(x,t)$ from the quenched one-soliton profile (\ref{psi0}) with $\eta=2$, $\varkappa = 1$, $u = 0$, $\varphi = 0$ calculated from the exact expressions Eqs. (\ref{psisol})-(\ref{psisol1})}
\label{NS_Att_2}
\end{figure}
\end{center}

\begin{center}
\begin{figure}[th]
\includegraphics[width=1\textwidth]{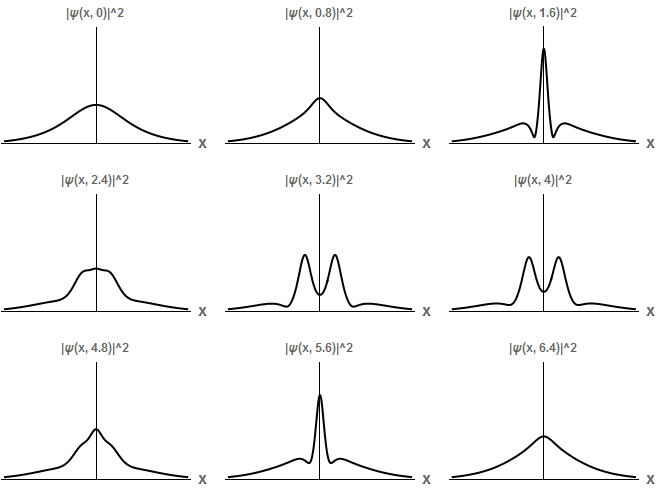}
\caption{Snapshots of the solution of the NLS equation $\psi(x,t)$ from the quenched one-soliton profile (\ref{psi0}) with $\eta=3$, $\varkappa = 1$, $u = 0$, $\varphi = 0$ calculated from the exact expressions Eqs. (\ref{psisol})-(\ref{psisol1})}
\label{NSAtt3}
\end{figure}
\end{center}

The corresponding auxiliary linear problem for equation (\ref{NLSa}) is
\be\label{NLSaF}
\frac{d F}{d x} =\left(
 \begin{array}{cc}
        \displaystyle\frac{\lambda}{2i} & i\sqrt{|\kappa|}\bar{\psi}(x) \\
        i\sqrt{|\kappa|}\psi(x) & -\displaystyle\frac{\lambda}{2i} \\
      \end{array}
    \right)F.
\ee
The Jost solutions $T_{\pm}(x)$ are defined by their asymptotic
\be
T_{\pm}(x) \to e^{-i\sigma_3\lambda x/2},\,\,\,\,\,\, x\to \infty.
\ee
Let us consider the following initial condition for the field $\psi(x)$
\be\label{psi0}
\psi(x) = \frac{e^{i\varphi}}{\sqrt{|\kappa|}} \frac{e^{iux}}{\cosh(x/\eta)}.
\ee
If $\eta=1$, the whole profile moves with constant velocity without changing its shape. 
For generic $\eta$, according to our strategy, we must first solve the direct scattering problem (\ref{NLSaF}) on the solution (\ref{psi0}), which can be done by reducing (\ref{NLSaF}) to the Fuchs form. To do this we perform a gauge transformation 
\be
F\to e^{ix\lambda/2}\left(
 \begin{array}{cc}
        0 & e^{-x/\eta - i u x} \\
        1 & 0 \\
      \end{array}
    \right)F
\ee
and use a coordinate transformation $z = (1+\tanh(x/\eta))/2$ to present our equations in the form of Eq. (\ref{1.1}) with matrices 
\be
A_0 = \left(
\begin{array}{cc}
0 & i\eta e^{i\varphi}  \\
0 & \frac{1}{2}+i(u-\lambda)\eta/2
\end{array}
\right),\,\,\,\,\,\,
A_1 = \left(
\begin{array}{cc}
0 & 0  \\
i\eta e^{-i\varphi} & \frac{1}{2}+i(u-\lambda)\eta/2
\end{array}
\right).
\ee
By matching the general asymptotic with Eq. (\ref{Jost}) we recover matrices that specify the Jost solutions, namely
\be
C_0=C_1 = \sigma_1\,.
\ee
The general transfer matrix that comes from (\ref{transfer}) has a form of  
\be\label{TTT}
T(\lambda) = \left(
                                                                                            \begin{array}{cc}
                                                                                              a(\lambda) & -\bar{b}(\lambda) \\
                                                                                              b(\lambda) & \bar{a}(\lambda) \\
                                                                                            \end{array}
                                                                                          \right)
\ee
with 
\be
a(\lambda) = \frac{\Gamma\left(\frac{1}{2}+\frac{i(u-\lambda)\eta}{2}\right)^2}{\Gamma\left(\frac{1}{2}+\frac{i(u-\lambda)\eta}{2}-\eta\right)\Gamma\left(\frac{1}{2}+\frac{i(u-\lambda)\eta}{2}+\eta\right)},
\ee
\be
b(\lambda) = ie^{i\varphi} \frac{\sin(\pi \eta )}{\cosh(\pi \eta(u-\lambda)/2)}.
\ee
The solitonic solutions ($b(\lambda)=0$) correspond to 
\be
\eta = n,\,\,\,\,\,\,\,\, n \in \mathds{Z}.
\ee
Further, without any loss of generality we can assume that $n>0$. The diagonal components then become equal to
\be\label{na}
a(\lambda) = \prod \limits_{k=1}^n \frac{\lambda-u -(2k-1)i/\eta}{\lambda-u +(2k-1)i/\eta}.
\ee
For this form of diagonal components, the exact solution given by the IST can be written as \cite{Faddeev}
\be\label{psisol}
\psi(x,t) = -\frac{\exp\left(i\varphi+iux-iut^2\right) }{\sqrt{|\kappa|}} \frac{\det M_1(x,t)}{\det M(x,t)}
\ee
where $M(x,t)$ is an $n\times n$ matrix with elements
\be
M_{jk}(x,t) = \frac{i\eta}{2}\frac{1+\bar{\gamma}_j(x,t)\gamma_k(x,t)}{j+k-1},
\ee
\be
M_1 =
\begin{blockarray}{ccc}
 \begin{block}{(ccc)}
     &   &    \bar{\gamma}_1(x,t)        \\  
   & \Big{M} & \vdots  \\  
     &  &\bar{\gamma}_n(x,t)                 \\
   1 & \cdots 1 &  0              \\ 
 \end{block}
 \end{blockarray}
\ee
and $\gamma_j$ are given by
\be
\gamma_j(x,t) = \gamma_j\exp\left(-\frac{2j-1}{\eta} (x-2ut)-it\frac{(2j-1)^2}{\eta^2} \right).
\ee
The constants $\gamma_j$ should be determined from the initial conditions ($\ref{psi0}$). At $t=0$ we can introduce 
variable $Z = e^{-x/\eta}$, such that $\gamma_j(x,0)= \gamma_j Z^{2j-1}$. Therefore, the initial condition takes a form of 
\be\nonumber
\frac{\det M_1(x,0)}{\det M(x,0)} = \frac{2Z}{Z^2+1}\,,
\ee
which is just a restriction on the rational functions of $Z$, from which all the coefficients $\gamma_j$ can be determined.
In our case this can be done explicitly and the answer is 
\be\label{psisol1}
\gamma_j = i (-1)^{n+j-1}\,.
\ee
We see that unlike to the KdV case, the initial profile does not split into  
separate one-solitonic solutions, but rather all quench-produced solitons move with the same velocity and the enveloping shape is oscillating.  
The reason for this is that the velocity of the NLS soliton is not connected with its width and/or amplitude. 
Therefore, it is sufficient to consider the case of zero initial velocity $u=0$, and the generic case can be obtain 
by the Galilean transformation to the frame moving with velocity $2u$.
The exemplary post-quench dynamics for $\eta=2$ and $\eta=3$ is shown in 
Fig. \ref{NS_Att_2} 
 and Fig. \ref{NSAtt3} respectively.

The integrals of motion are given by an 
integral of the local densities $w_n(x)$
\be
I_n = \kappa\int\limits_{-\infty}^\infty dx \bar{\psi}(x)w_n(x)\,,
\ee
which are defined recursively
\be
w_{n+1}(x) = - i\frac{dw_n(x)}{dx} + \kappa \bar{\psi}(x)\sum\limits_{k=1}^{n-1}w_k(x)w_{n-k}(x),
\ee
\be
w_1 = \psi(x).
\ee
The first three integrals of motion computed from the profile (\ref{psi0}), are
\be
I_1 =  -2\eta,\,\,\,\,\,\,
I_2 = -2u\eta,\,\,\,\,\,\,I_3 = -\frac{2}{3 \eta }-2 u^2 \eta + \frac{4\eta}{3  }.
\ee
An other way to deduce those integrals is to use the asymptotic expansion for $\ln a(\lambda)$ \cite{Faddeev}
\be
\ln a(\lambda)=i\sum\limits_{j=1}^\infty \frac{I_j}{\lambda^j} + O(|\lambda|^{-\infty}),
\ee 
which produces the form for the $I_j$ integral of motion
\be
I_j =  \sum\limits_{k=1}^{\eta} \frac{i}{j}\left[\left(u+\frac{2k-1}{\eta}i \right)^j-
\left(u-\frac{2k-1}{\eta}i \right)^j \right].
\ee
Similar, to the KdV case we can introduce a simple generating function for these integrals of motion, that after some resummations 
can be expressed in terms of elementary functions, namely
\be
S(t) = \sum\limits_{j=1}^\infty \frac{I_j t^j}{(j-1)!} = -2\frac{e^{ut}\sin^2t}{\sin (t/\eta)}.
\ee
This function is a bit simpler than $\ln a(\lambda)$ and the values for the integrals of motion are valid for any $\eta$.

\section{Sine-Gordon equation}
\label{SG}

The sine-Gordon equation (SG)
\be\label{SG1}
\frac{\partial^2 \varphi}{\partial t^2} - \frac{\partial^2 \varphi}{\partial x^2} +  \frac{m^2}{\beta} \sin \beta \varphi = 0
\ee
has numerous applications in various branches of physics, starting from  
dislocations in solids to phase dynamics in long Josephson junctions. The equivalence with a period $2\pi/\beta$ is assumed  
for a real function $\varphi(x,t)$ along with the rapidly decreasing boundary conditions $\lim\limits_{|x|\to \infty} \varphi(x) = 0({\rm mod}\, 2\pi/\beta)$. 
The system is characterized by the topological charge 
\be
Q = \frac{\beta}{2\pi} \int\limits_{-\infty}^\infty \frac{\partial \varphi(x)}{\partial x} dx\,,
\ee
which being integer imposes certain restrictions to the possible quench. 

To realize quench program outlined in the previous chapters it is useful to introduce light-cone coordinates
\be
u =  \frac{t+x}{2}, \quad\quad v =  \frac{t-x}{2},
\ee
so Eq. (\ref{SG1}) transforms into
\be\label{SG2}
\frac{\partial^2 \varphi}{\partial u \partial v}  +  \frac{m^2}{\beta} \sin \beta \varphi = 0\,.
\ee
In this form SG equation appeared in the middle of the 19th century in the context of geometry problems of surfaces with constant negative curvature. 

The one-soliton solution in $(x,t)$ coordinates has the form 
\be
\varphi(x,t) = \frac{4Q}{\beta} \arctan \exp \left(\frac{m(x-ct)}{\sqrt{1-c^2}}\right)
\ee
where $|c|<1$ is the velocity of the soliton and topological number $Q$ for this solution acquires only two possible values $Q = \pm 1$.
The same solution in the light-cone coordinates is 
\be 
\varphi (u,v) = \frac{4Q}{\beta} \arctan \exp \left(m \varkappa  u-\frac{ m v}{\varkappa}\right),\,\,\,\,\,\, \varkappa = \sqrt{\frac{1-c}{1+c}}.
\ee
We see that the width of the soliton is not connected with its amplitude as it was in all previous examples. 
Therefore, in this section we consider a scaling of the amplitude, which physically means that we perform quench in a parameter $\beta \to \beta /\eta$, while fixing $Q=+1$ and $m=1$.  
This way, our initial condition reads as
\be\label{InitialSG} 
\varphi_\eta (u,v=0) = \frac{4\eta }{\beta} \arctan e^{\varkappa u} .
\ee
The corresponding auxiliary linear problem for equation (\ref{SG2}) is
\be\label{SGF}
\frac{d F}{d u} =\left(
 \begin{array}{cc}
        \displaystyle\frac{\beta}{4i}\frac{\partial \varphi_\eta}{\partial u} & \displaystyle-\frac{\lambda}{2}e^{i\beta \varphi_\eta/2} \\
        \displaystyle\frac{\lambda}{2}e^{-i\beta \varphi_\eta/2} & -\displaystyle\frac{\beta}{4i}\frac{\partial \varphi_\eta}{\partial u} \\
      \end{array}
    \right)F.
\ee 
One can consider transformation to the light-cone coordinates as a gauge transformation that involves 
not only matrix $U$ but also a matrix $V$ from the linear system Eqs. (\ref{U}),(\ref{V}). 

Performing additional gauge transformation
\be
F \to 
\left(
 \begin{array}{cc}
        \displaystyle 0 & \displaystyle ie^{i\beta \varphi_\eta/4} \\
        \displaystyle \displaystyle ie^{-i\beta \varphi_\eta/4} & 0 \\
      \end{array}
    \right)
F
\ee 
we reduce system (\ref{SGF}) to 
\be
\frac{d F}{d u} =\left(
 \begin{array}{cc}
        \displaystyle\frac{\lambda}{2i} & \displaystyle\frac{\beta}{2}\frac{\partial \varphi_\eta}{\partial u} \\
        -\displaystyle\frac{\beta}{2}\frac{\partial \varphi_\eta}{\partial u} & - \displaystyle\frac{\lambda}{2i} \\
      \end{array}
    \right)F = \left(
 \begin{array}{cc}
        \displaystyle\frac{\lambda}{2i} & \displaystyle\frac{\eta \varkappa}{\cosh(u\varkappa)} \\
        -\displaystyle\frac{\eta\varkappa}{\cosh(u\varkappa)} & - \displaystyle\frac{\lambda}{2i} \\
      \end{array}
    \right)F.
\ee
We see that this scattering problem is analogous to the attractive NLS (\ref{NLSaF}) with some special choices of the initial condition (\ref{psi0}) and an additional rescaling of the spatial coordinate. Therefore, the transfer matrix 
has the form (\ref{TTT})
with 
\be
a(\lambda) = \displaystyle\frac{\Gamma\left(\frac{1}{2}-\frac{i\lambda }{2\varkappa}\right)^2}{\Gamma\left(\frac{1}{2}-\frac{i\lambda }{2\varkappa}-\eta\right)\Gamma\left(\frac{1}{2}-\frac{i\lambda }{2\varkappa}+\eta\right)},
\ee
\be
b(\lambda) =-\frac{\sin(\pi \eta )}{\cosh[\pi\lambda /(2\varkappa)+i\pi\eta]}.
\ee
So again for integer $\eta$ we have reflectionless potential and the coefficient $a(\lambda)$ reads as ($\eta=n\in \mathds{Z}_{>0}$)
\be\label{aSG}
a(\lambda) = \prod\limits_{k=1}^n \frac{\lambda - i \varkappa (2k-1)}{\lambda + i \varkappa (2k-1)}.
\ee
Such a form corresponds to the $n$-solitonic solution. Here we would like to stress that this is the only possible
choice of $\lambda$ to satisfy periodic boundary conditions. 

\begin{center}
\begin{figure}[h]
\includegraphics[width=1\textwidth]{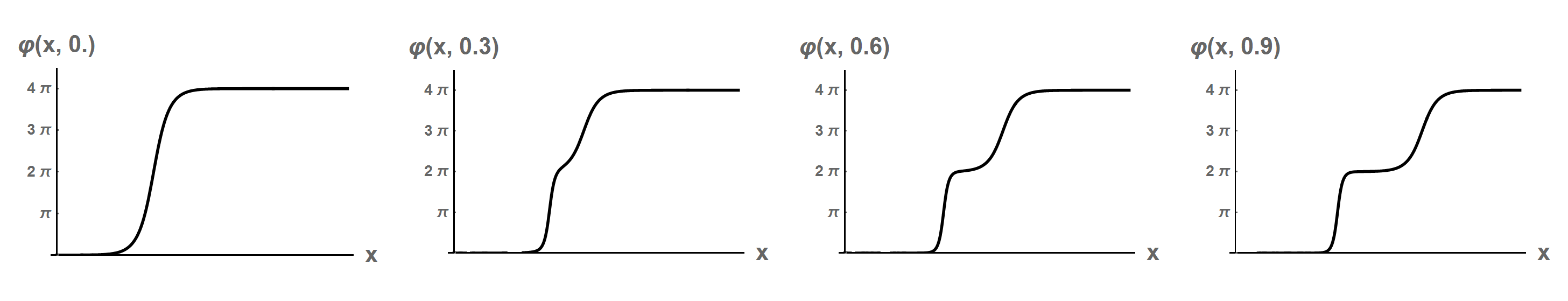}
\caption{Snapshots of the solution of the SG equation $\varphi(x,t)$ from the quenched one-soliton profile (\ref{InitialSG}) with $\eta=2$, $\varkappa = 0.1$, $\beta= 1$ calculated from the exact expressions Eqs. (\ref{SGsol})-(\ref{SGsol1})}
\label{SGpic}
\end{figure}
\end{center}

The explicit solution of the IST is given by
\be\label{SGsol}
\varphi = \frac{2i}{\beta} \ln\frac{\det(1+V)}{\det(1-V)}
\ee
where
\be
V_{jk} = \frac{\beta_k}{\varkappa_j+\varkappa_k} \exp\left(\varkappa_k u - \frac{v}{\varkappa_k}\right),\;\; \varkappa_k=\varkappa(2k-1),
\ee
\be\label{SGsol1}
\beta_k = \frac{-2\varkappa i}{((k-1)!)^2 }\frac{(n+k-1)!}{(n-k)!}.
\ee
The time evolution's snapshots for $n=2$ are shown in Fig. \ref{SGpic}.

The local integrals of motion can be easily found via identification with NLS model. Namely, they can be calculated by the recurrence 
procedure
\be
I_n=i\beta^2/4\int\limits_{-\infty}^{+\infty}du,\; w_n(u)\frac{d\varphi(u)}{du},
\ee 
\be
w_{n+1}(u)=-i\frac{d w_n(u)}{du}+i\frac{\beta^2}{4}\frac{\varphi(u)}{du}\sum\limits_{k=1}^{n-1}w_{k}(u)w_{n-k}(u),\;\;w_1(u)=i\frac{\varphi(u)}{du}.
\ee
All even densities are total derivatives so we consider only the odd integrals of motion.
They can be explicitly computed from the initial profile (\ref{InitialSG}) and the first three integrals of motion are
\be 
\fl \quad\quad 	I_1=-2\eta^2\varkappa,\;\;\; I_3=\frac{2}{3}(-\eta^2+2\eta^4)\varkappa^3,\;\;\; I_5=-\frac{2}{15}(7\eta^2-20\eta^4+16\eta^6)\varkappa^5.
\ee
Another way to compute these integrals is through the asymptotic expansion of $\ln a(\lambda)$, namely
\be
\ln a(\lambda)=i\sum\limits_{j=0}^{\infty}\lambda^{-j}I_{j},
\ee
which gives, for an integer $\eta$,
\be
I_{j}=i \sum\limits_{k=1}^{\eta}\frac{(i\varkappa(2k-1))^{j}}{j}(1-(-1)^{j}).
\ee
As in the previous cases we can rearrange the generating series for the integrals of motion to present them in terms of the elementary functions, namely
\be
S(t)=\sum\limits_{j=1}^{\infty}\frac{I_j t^j}{(j-1)!}=\frac{\cos(2\eta t\varkappa)-1}{\sin(t\varkappa)}\,.
\ee
Again, we would like to emphasize, that even though this formula has been obtained for a solitonic $\eta$ --- it is also correct for in generic case.

\section{Non-linear Schr\"odinger equation. Repulsive case at finite density}

\label{NLS}

\begin{center}
\begin{figure}[h]
\includegraphics[width=1\textwidth]{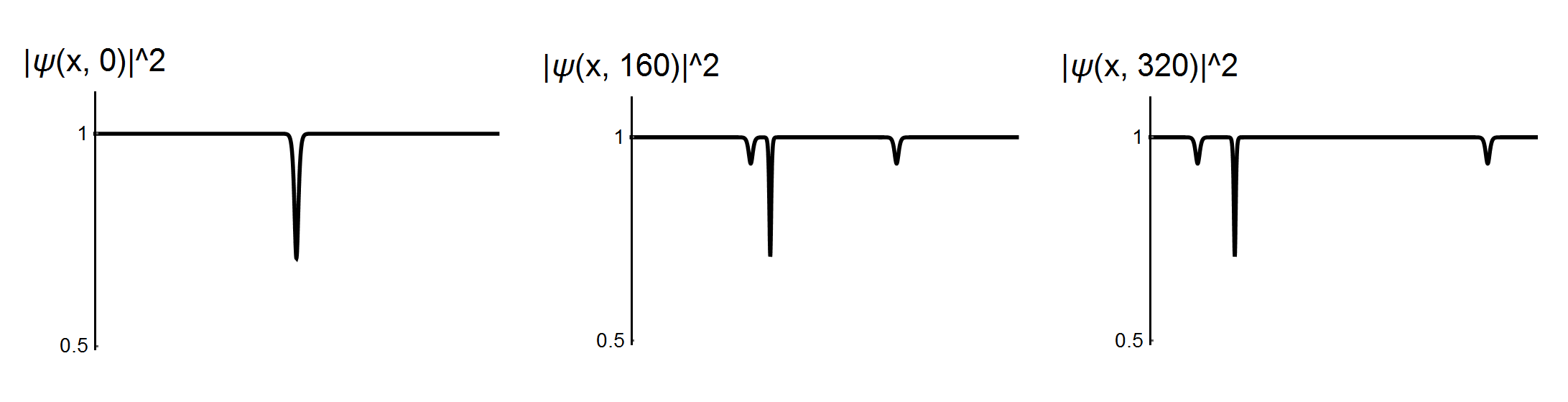}
\caption{Snapshots of the solution of the repulsive NLS equation $\psi(x,t)$ from the quenched one-soliton profile (\ref{ppsi0}) with $n=2$, $\theta=\pi/2$ calculated from the exact expressions Eqs. (\ref{rsol})-(\ref{rsol1}).}
\label{reprep2}
\end{figure}
\end{center}

\begin{center}
\begin{figure}[h]
\includegraphics[width=1\textwidth]{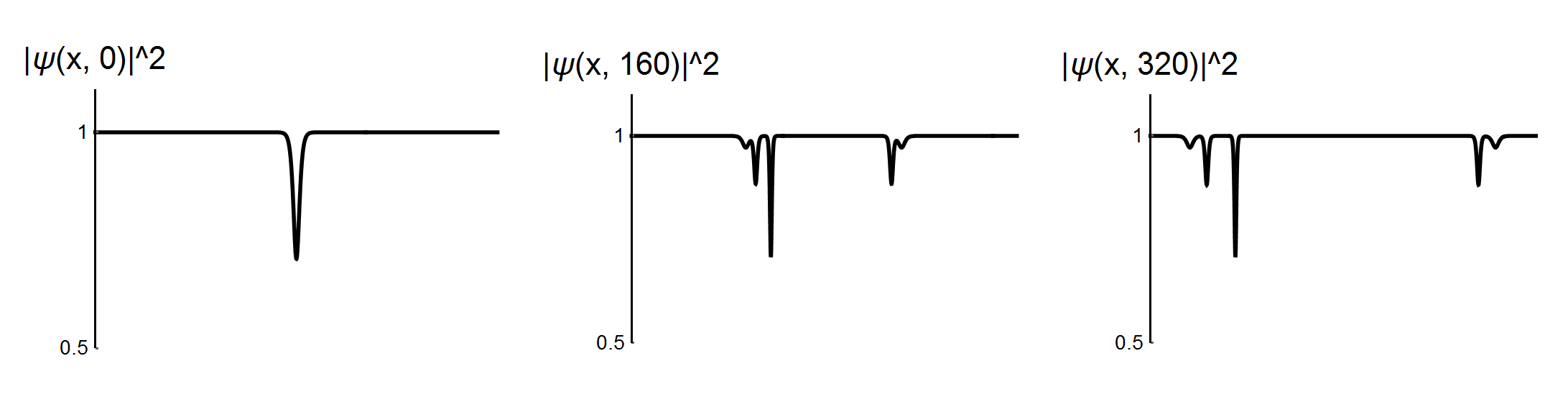}
\caption{Snapshots of the solution of the repulsive NLS equation $\psi(x,t)$ from the quenched one-soliton profile (\ref{ppsi0})  with $n=3$, $\theta=\pi/2$ calculated from the exact expressions Eqs. (\ref{rsol})-(\ref{rsol1}).}
\label{reprep3}
\end{figure}
\end{center}

Finally, we present the solution for the NLS equation in the repulsive case at finite density. Main results of this section were obtained earlier in Ref. \cite{PhysRevA.91.031605}, and here we give more detailed and coherent presentation based on the general scheme. 
NLS equation at finite density is given by
\be\label{nsLEq}
i\frac{\partial \psi}{\partial t} = -\frac{\partial^2\psi}{\partial x^2} + 2\kappa (|\psi|^2-1)\psi\,,
\ee
where we assume that $\kappa>0$ so for our convenience and without loss of generality we can put $\kappa= 1/4$.
The asymptotic conditions correspond to finite density $|\psi(x\to \pm \infty)| \to 1$ and the phase difference 
\be
\psi(x)\to e^{i\theta},\,\,\,\,\,\,\, x \to +\infty,\,\,\,\,\,\,\,
\psi(x)\to 1 ,\,\,\,\,\,\,\, x \to -\infty\,.
\ee
The corresponding linear problem reads as
\be\label{nsL}
\frac{d F}{d x} =\left(
 \begin{array}{cc}
        \displaystyle\frac{\lambda}{2i} & \displaystyle\frac{\bar{\psi}(x)}{2} \\
        \displaystyle\frac{\psi(x) }{2}& -\displaystyle\frac{\lambda}{2i} \\
      \end{array}
    \right)F.
\ee
The Jost solutions are determined by their asymptotic 
\be
T_-(x\to-\infty) = S e^{-ik x \sigma_3/2},\,\,\,\,\,T_+(x\to-\infty) = e^{-i\theta \sigma_3/2}S e^{-ik x \sigma_3/2}
\ee
where 
\be
S = \left(
 \begin{array}{cc}
        1 & i(k-\lambda) \\
        i(\lambda-k) & 1 \\	
      \end{array}
    \right),\,\,\,\,\,\, k = \sqrt{\lambda^2-1}.
\ee
As the initial value we consider a scaled one-soliton solution
\be\label{ppsi0}
\psi(x) = \frac{1 +  e^{i\theta} e^{2x/\eta}}{1 +  e^{2x/\eta}}.
\ee 
To solve the scattering problem (\ref{nsL}) we perform a gauge transformation 
$
F \to e^{-i k x/2}F
$
and 
introduce the
variable $z=(1+\tanh(x/\eta))/2$, such that (\ref{nsL}) takes the Fuchs form with matrices
\be
A_0 = \frac{i\eta}{2}\left(
 \begin{array}{cc}
        0 & 1 \\
        1  & \lambda \\
      \end{array}
    \right),\,\,\,\,\,
A_1 = \frac{i \eta}{2}\left(
 \begin{array}{cc}
        0 & e^{-i\theta} \\
        e^{i\theta}  & \lambda \\
      \end{array}
    \right).
\ee
The corresponding matching matrices are 
\be
C_0 = \left(
 \begin{array}{cc}
        1 & 0 \\
        0  & 1 \\
      \end{array}
    \right), \,\,\,\,\, C_1 = \left(
 \begin{array}{cc}
        e^{-i\theta/2} & 0 \\
        0  & e^{i\theta/2} \\
      \end{array}
    \right).
\ee 
The transfer matrix (\ref{transfer}) has a form of
\be
T= \left(
 \begin{array}{cc}
        a(\lambda) & \bar{b}(\lambda) \\
        b(\lambda)  & \bar{a}(\lambda) \\
      \end{array}
    \right)
\ee
with
\be
\fl \quad\quad a(\lambda) = \frac{\Gamma(-ik\eta/2)^2}{\Gamma(-ik\eta/2-\eta \sin \frac{\theta}{2})\Gamma(-ik\eta/2+\eta \sin \frac{\theta}{2})}\frac{k}{k\cos(\theta/2)-i\lambda\sin(\theta/2)},
\ee
\be
\fl \quad\quad b(\lambda) = \frac{ \sin\left[ \frac{\pi  \eta  }{2}  \sin \left(\frac{\theta }{2}\right)\right]}{\sinh\left[\frac{\pi  k \eta }{2}\right]}.
\ee
The solitionic solutions correspond to \be\eta = 2n/\sin(\theta/2)\label{etta},\ee for integer $n$, which, further, without loss of the generality we can assume to be 
positive. Then the form of the $a(\lambda)$ reads as
\be\label{nnaa}
a(\lambda) = \frac{ik+\sin\frac{\theta}{2}}{ik \cos\frac{\theta}{2}+\lambda\sin\frac{\theta}{2}} \prod\limits_{s=1}^{n-1} \frac{k-2is/\eta}{k+2is/\eta}.
\ee
To present this expression in the usual solitonic form we introduce $y=k+\lambda$, such that $k= (y-y^{-1})/2$ and $\lambda = (y+y^{-1})/2$. Then Eq. (\ref{nnaa}) can be written as
\be
a(\lambda)  = e^{i\theta/2} \frac{y+e^{-i\theta/2}}{y+e^{i\theta/2}}\prod\limits_{s=1}^{n-1} \frac{y+e^{-i\theta_s^+/2}}{y+e^{i\theta_s^+/2}}\frac{y+e^{-i\theta_s^-/2}}{y+e^{i\theta_s^-/2}}\,,
\ee
which now constitute the usual solitonic form \cite{Faddeev} of $2n-1$ solitons with a set of the parameters $\Theta$, 
\be
\Theta=\{\theta_1,\theta_{2},\dots \theta_{2n-1}\} = \{\theta, \theta_1^+,\dots, \theta^+_{n-1},\theta_1^-,\dots, \theta^-_{n-1}\}
\ee
where
\be
\theta_s^+ = 2\arcsin \frac{s}{n}\sin\frac{\theta}{2},\quad
\theta_s^- = 2\pi-\theta_s^+.
\ee
The corresponding solution is given by the ratio of the determinants of two $(2n-1)\times (2n-1)$ matrices, namely
\be\label{rsol}
\Psi(x,t) = \frac{{\rm det}(1+\tilde{A})}{{\rm det}(1+ A)}
\ee
where
\be
\fl \quad A_{jk} = \frac{2i\sqrt{\beta_j\beta_k}}{e^{i\theta_k/2}-e^{-i\theta_j/2}}\exp\left(
\frac{x}{2}\left(\nu_j+\nu_k\right) -\frac{q_j+q_k}{4}
\right),\,\,\,\,\,\,\,\, \tilde{A}_{jk} = A_{jk}\exp\left(i\frac{\theta_j+\theta_k}{2}\right)
\ee
and
\be\label{rsol1}
\fl \quad
\nu_j = \sin\frac{\theta_j}{2},\,\,\,\, q_j =- t \sin\theta_j,\,\,\,\,\,\ \beta_j = \left|\sin\frac{\theta_j}{2}\prod\limits_{k\neq j} \frac{\sin\frac{\theta_j+\theta_k}{4}}{\sin\frac{\theta_j-\theta_k}{4}}\right|,\,\,\, j=1,2,\dots 2n-1.
\ee
Another useful and more explicit combinatorial form for the coefficients $\beta_j$ is
\be
 \{\beta_1,\beta_{2},\dots \beta_{2n-1}\} = \{\beta_0, \beta_1^+,\dots, \beta^+_{n-1},\beta_1^-,\dots, \beta^-_{n-1}\}.
\ee
\be
\beta_0 = \left|\sin \frac{\theta}{2} \right| \frac{\Gamma(2n)}{\Gamma(n+1)\Gamma(n)},
\ee
\be
\beta^{\pm}_j = \frac{j|\sin \theta/2|}{\sqrt{n^2-(j \sin \theta/2)^2}}\frac{\Gamma(n+j)}{2\Gamma(j+1)^2\Gamma(n-j+1)} \left|\frac{\sin\frac{\theta+\theta^\pm_j}{4}}{\sin\frac{\theta-\theta^\pm_j}{4}} \right|.
\ee
Alternatively we can write the solution (\ref{rsol}) in the following form
\be\label{rsol2}
\fl \quad\Psi(x,t) = \frac{1+ \sum\limits_{1\le j_1<\dots j_l\le 2n-1}e^{(\nu_{j_1}x-q_{j_1}+i\theta_{j_1})+\dots+(\nu_{j_k}x-q_{j_k}+i\theta_{j_k})} \prod\limits_{j=j_1,\dots j_l}\prod\limits_{k\neq j_1,\dots j_l}\left|\frac{\sin\frac{\theta_{j}+\theta_k}{4}}{\sin\frac{\theta_{j}-\theta_k}{4}}\right|
}{1+ \sum\limits_{1\le j_1<\dots j_l\le 2n-1}e^{(\nu_{j_1}x-q_{j_1})+\dots+(\nu_{j_k}x-q_{j_k})} \prod\limits_{j=j_1,\dots j_l}\prod\limits_{k\neq j_1,\dots j_l}\left|\frac{\sin\frac{\theta_{j}+\theta_k}{4}}{\sin\frac{\theta_{j}-\theta_k}{4}}\right|}.
\ee
Expressions (\ref{rsol}) or (\ref{rsol2}) remain the solutions for Eq. (\ref{nsLEq}) even if we add arbitrary real numbers to the $q_j$. 
These numbers play a role of the initial positions of solitons and in our case we have chosen them to be zero to match the initial distribution (\ref{ppsi0}).
The exemplary post-quench dynamics is shown in Fig. \ref{reprep2} and Fig. \ref{reprep3} respectively.

The polynomial integrals of motion can be found in a similar manner as in previous sections. The first three integrals of motion evaluated from the initial field configuration (\ref{ppsi0}) are 
\be
\fl \quad I_1 = \int\limits_{-\infty}^\infty dx (|\Psi(x)|^2-1)  = -\eta (1-\cos\theta),
\ee
\be
\fl \quad I_2 = \int\limits_{-\infty}^\infty \frac{dx}{2i} \left(
\frac{\partial \Psi (x)}{\partial x} \bar{\Psi}(x) - 
\frac{\partial \bar{\Psi} (x)}{\partial x} \Psi(x)
\right)   = \sin \theta,
\ee
\be
\fl \quad I_3 = \int\limits_{-\infty}^\infty dx \left(\left|\frac{\partial \Psi (x)}{\partial x}\right|^2+\frac{(1-|\Psi(x)|^2)^2}{4}\right)  = 
\frac{\sin ^2\left(\theta/2\right) \left(\eta ^2 (1-\cos (\theta ))+8\right)}{6 \eta }.
\ee
Another way to generate these integrals of motion is to use the asymptotic expansion of the $\ln a(\lambda)e^{-i\theta/2}$, namely,
\be
-i\ln a(\lambda) e^{-i\theta/2} = \frac{1}{4} \sum\limits_{s=1}^\infty \frac{I_s}{k^s}\,.
\ee
$a(\lambda)$ written in a form of Eq. (\ref{nnaa}) is obviously factorizable to the quench dependent and quench independent parts. The asymptotic expansion of the latter produces
quench dependent integrals of motion, namely
\be
I^\eta_j = 4i \sum\limits_{s=1}^{n-1} \left(\frac{2i s}{
\eta}\right)^j \frac{1-(-1)^j}{j}.
\ee
The generating function for these integrals can be computed in a way similar to those that we have used in previous cases. We have 
\be
S_\eta(t) = \sum\limits_{j=1}^\infty \frac{t^jI^\eta_j }{(j-1)!} = 4\frac{\cos\left(t \left(\frac{1}{\eta }-\sin\frac{\theta }{2}\right)\right) -\cos\left(t/\eta \right)}{\sin(t/\eta )}.
\ee
Note that this function is identically zero in the unquenched situation $1/\eta =1/2 \sin(\theta/2)$ ($n=1$ in Eq. (\ref{etta})). 
Therefore the total integrals of motion differ from the quenched ones only by adding one-soliton integrals of motion. One can find their explicit form in Ref. \cite{Faddeev}.

\section{Concluding remarks}

\label{Conclusions}

To conclude, we have considered
theoretical problem of quenching one-soliton solutions in the classical integrable equations.
We have noticed that the direct scattering problem that corresponds to the potential of a quenched soliton profile after gauge and coordinate transformations 
can be reduced to the Fuchs differential equation with three regular singular point. Generic solutions of these equations can be expressed in terms of hypergeometric functions, which allows us
to find an explicit form for the transfer matrix. Based on this result we have explored case of KdV, NLS and SG equations. In each case we have found conditions on the quench such that obtained
transfer matrix is diagonal (the potential to be reflectionless). The post-quench evolution of the initial profile, in this case, contains only solitons and parameters of these solitons can be easily found.   

We see, in particular, that when the width of the KdV or repulsive NLS soliton is quenched in the right way, it splits into solitons 
that travel with different velocities and after some time become manifestly distinct. The reason for that is relation between the width, amplitude and velocity of such type of solitons.
This is not the case for the attractive NLS soliton, for which geometric sizes and speed of propagation are independent.
Therefore after quenching the width of a soliton, no real splitting occurs but rather the enveloping shape of the post-quenched solitons  
exhibits oscillating behaviour. For SG soliton the width of the soliton is not related to its amplitude, and we find that 
the splitting occurs when the amplitude is quenched appropriately. After such a quench initial one-soliton profile splits into solitons 
that become distinct after some time, this way, resembling quench in the KdV and repulsive NLS cases. When the quench parameter does not satisfy the
quench condition then in addition to the solitons the radiation part is present, which has a form of small oscillating ripples. 

It is interesting to note that the quench condition for KdV equation involves triangular numbers. The same numbers appears when 
one considers pole dynamics in the complex plane of some special KdV solutions \cite{Kruskal,Thickstun1976335,AiraultMcKeanMoser,IMKrichiver,gesztesy1995,gesztesy1996,DeconinckSegur}.
Also the same numbers appear in algebro-geometrical potentials that correspond to "singular solitons" \cite{GrinevichNovikov}.   
We hope to elaborate more on this subject in future.

\section*{Acknowledgements}  
We are grateful to P. Gavrylenko, Yu. Zhuravlev and N. Iorgov for fruitful discussions and useful remarks. 
We also thank Mikhail B. Zvonarev, Oleg Lychkovskiy and especially Jimmy A. Hutasoit for careful reading of the
manuscript. This work is part of the Delta ITP consortium, a program of the Netherlands Organisation for Scientific Research (NWO) that is funded by the Dutch Ministry of Education, Culture and Science (OCW). 
The work of M. Semenyakin was partially supported by Project 1/30-2015  "Dynamics and topological structures in Bose-Einstein condensates of ultracold gases"  of the KNU  Branch Target Training at the NAS of Ukraine.

\newpage

\section*{References}

\bibliographystyle{iopart-num}
\bibliography{Quench3}

\end{document}